%% file: paper.tex
\documentclass[12pt]{iopart}
\usepackage{graphicx}
\usepackage{amsfonts}
\usepackage{amssymb}
\usepackage{mathptmx}
\usepackage{apjfonts}
\usepackage{natbib}

\input{defs}

\begin{document}

\bibliographystyle{jphysicsB}

\title[Accretion flows around coalescing SBHs]{Properties of accretion
flows around coalescing supermassive black holes}

\author{Tamara Bogdanovi\'c$^1$\footnote{Einstein Postdoctoral
Fellow}, Tanja Bode$^2$, Roland Haas$^2$, Pablo Laguna$^2$ and Deirdre
Shoemaker$^2$}

\address{$^1$Department of Astronomy, University of Maryland, College
Park, MD 20742-2421}

\address{$^2$Center for Relativistic Astrophysics, School of Physics,
Georgia Institute of Technology, Atlanta, GA 30332}

\ead{tamarab@astro.umd.edu}

\begin{abstract}
What are the properties of accretion flows in the vicinity of
coalescing supermassive black holes (SBHs)? The answer to this
question has direct implications for the feasibility of coincident
detections of electromagnetic (EM) and gravitational wave (GW) signals
from coalescences. Such detections are considered to be the next
observational grand challenge that will enable testing general
relativity in the strong, nonlinear regime and improve our
understanding of evolution and growth of these massive compact
objects. In this paper we review the properties of the environment of
coalescing binaries in the context of the {\it circumbinary disk} and
{\it hot, radiatively inefficient accretion flow} models and use them
to mark the extent of the parameter space spanned by this problem. We
report the results from an ongoing, general relativistic,
hydrodynamical study of the inspiral and merger of black holes, 
motivated by the latter scenario. We find that
correlated EM+GW oscillations can arise during the inspiral phase
followed by the gradual rise and subsequent drop-off in the light
curve at the time of coalescence. While there are indications that the
latter EM signature is a more robust one, a detection of either signal
coincidentally with GWs would be a convincing evidence for an
impending SBH binary coalescence. The observability of an EM
counterpart in the hot accretion flow scenario depends on the details
of a model. In the case of the most massive binaries observable by the
Laser Interferometer Space Antenna, upper limits on luminosity imply
that they may be identified by EM searches out to $z\approx
0.1-1$. However, given the radiatively inefficient nature of the gas
flow, we speculate that a majority of massive binaries may appear as
low luminosity AGN in the local universe.
\end{abstract}

\pacs{4.30.Tv, 04.70.-s, 95.30.Lz, 95.30.Sf, 98.62.Js, 98.62.Mw,
98.65.Fz}


\section{Introduction}

Supermassive black hole binaries (SBHBs) are one of the prime targets
for the future gravitational wave observatory, Laser Interferometer
Space Antenna (LISA). In anticipation of future gravitational wave
detections, a lot of effort has been directed towards the study and
characterization of these objects. Some of the most important open
questions pertain to the formation and cosmological evolution of the
supermassive black holes (SBHs), the rate of their coalescences and
associated observational signatures. All are intricately connected to
the properties of the environment in which the SBHs find themselves
during the cosmic time. An outstanding astrophysical question with
direct importance for the feasibility of coincident EM and GW
detections of coalescing SBHs is regarding the physical properties of
the gaseous environment surrounding a binary before and during
coalescence. Most of the information about these systems so far had to be
derived from theoretical studies and computational simulations since
finding them in EM searches proved to be a difficult task. Over the
past several years non-relativistic hydrodynamical simulations have
significantly contributed to our understanding of the evolution of BH
pairs and their host galaxies, both during and after the galactic
mergers \citep{kazantzidis05, an02, escala04, escala05, dotti07,
mayer07, colpi07, mm08, hayasaki08, cuadra09}. However, simulations
spanning the entire dynamical range, from galactic merger scales
($\sim 10^2 \,\hbox{kpc}$) to binary coalescences ($\ll
10^{-2}\,\hbox{pc}$), are still prohibitively computationally
expensive. As a consequence, non-relativistic simulations stop at
binary separations of $\sim 1\,\mathrm{pc}$ while fully general
relativistic simulations are possible only at separations of $\sim
10^{-5}\,\mathrm{pc}$. Hence, the properties and structure of
accretion flows around coalescing binaries remain uncertain.

In this paper, we discuss physically motivated scenarios for the SBHB
environments in centers of galaxies. We outline the properties of the
circumbinary disk and a hot and radiatively inefficient accretion flow
enveloping a binary and their implications for observations. Guided by
the radiatively inefficient accretion flow scenario we carried out
fully general relativistic, hydrodynamical study of the inspiral and
merger of equal-mass, spinning SBHBs. We describe the results from
this ongoing study which is a step towards understanding the how 
properties of a SBHB environment correlate with observational signatures.

\section{Physical conditions in the accretion flow}

\begin{figure}[ht]
\center{
\includegraphics[width=0.47\linewidth]{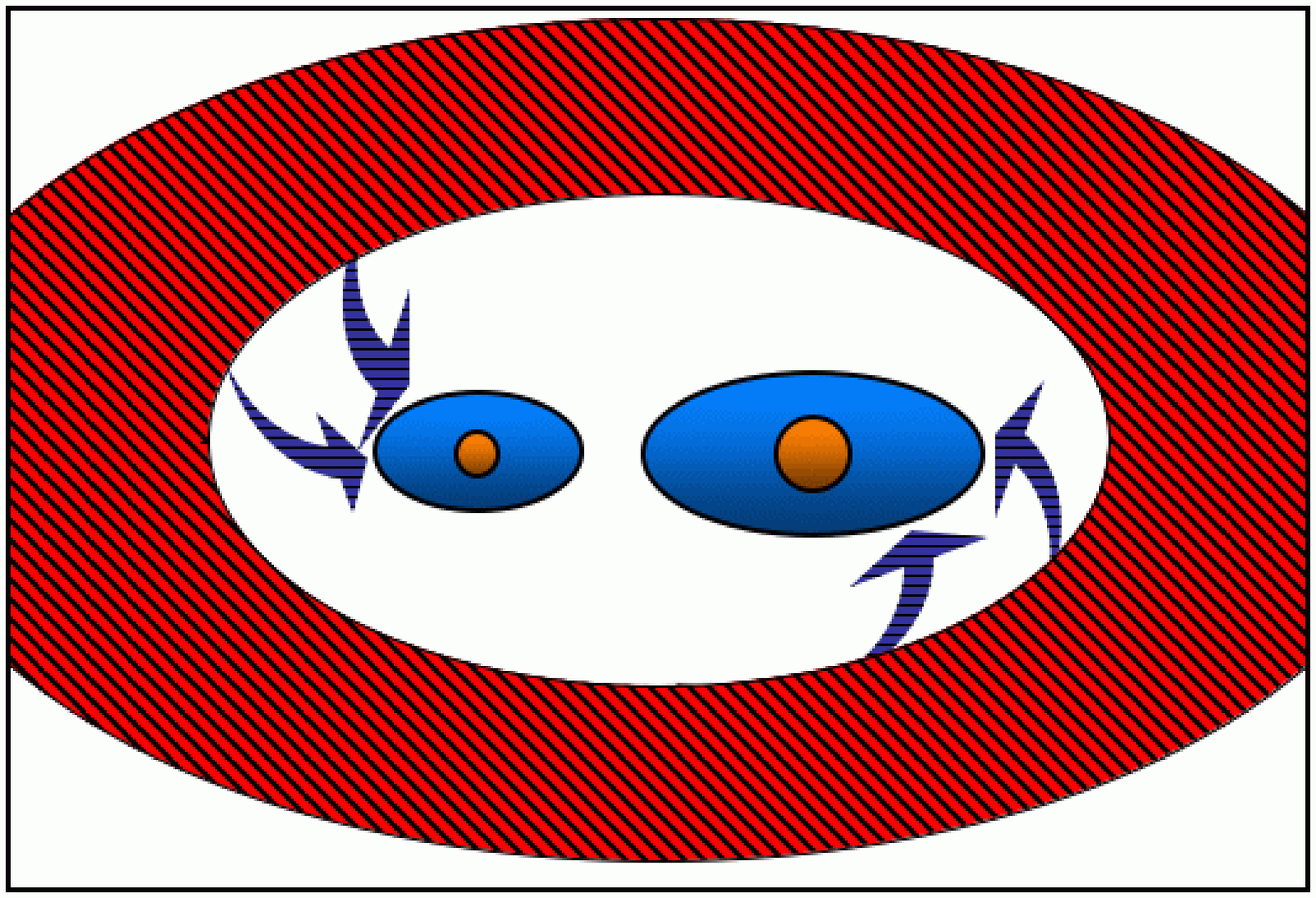}
\includegraphics[width=0.47\linewidth]{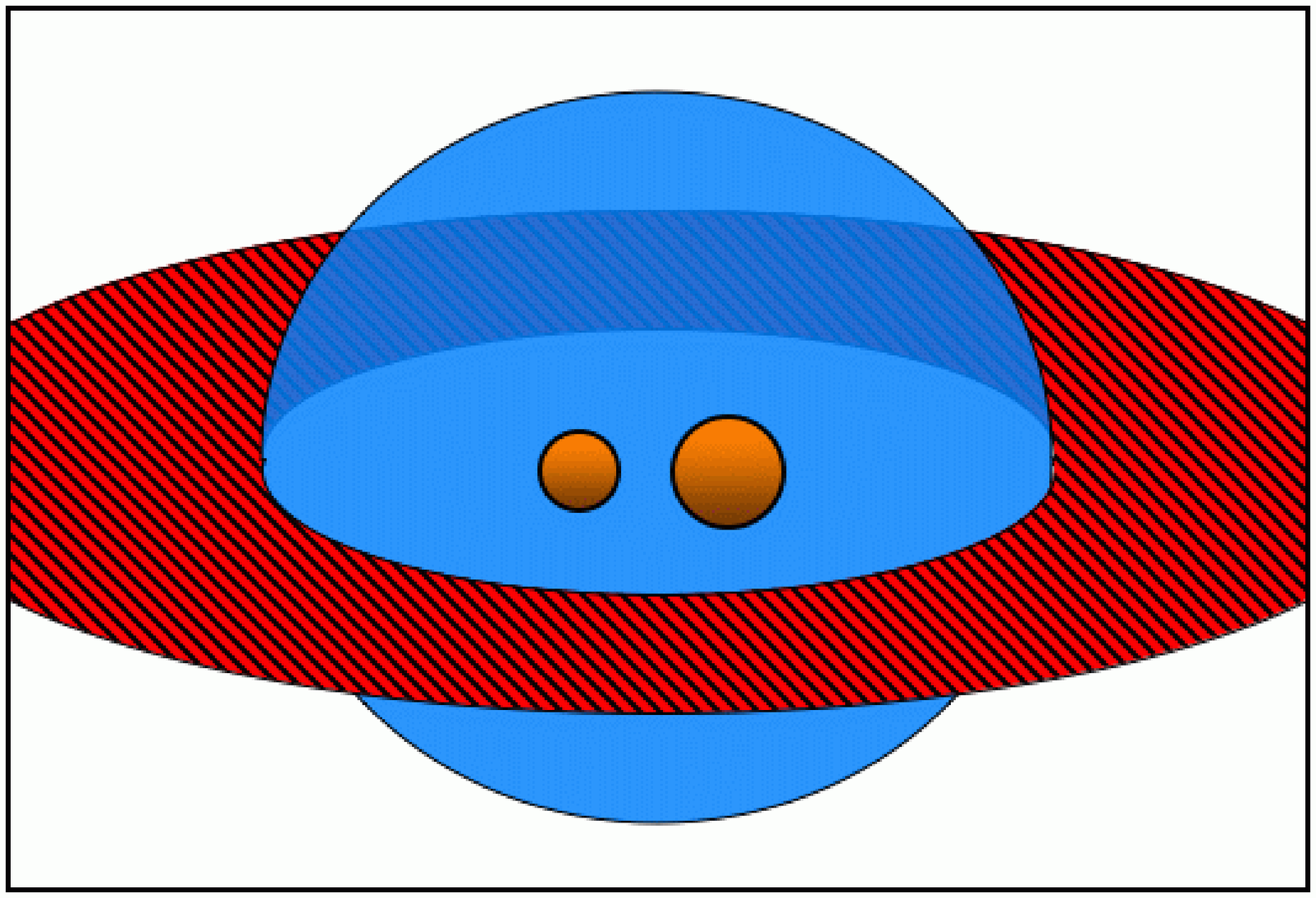}}
\caption{Illustration of the two accretion flow scenarios discussed in
this paper. {\it Left:} Circumbinary disk scenario in which binary
torques create a low density region in the center of the disk. The
accretion onto the binary members is mediated by their individual
circumblackhole disks as long as a balanced accretion rate can be
maintained from the circumbinary disk. {\it Right:} In radiatively
inefficient accretion flow scenario the SBHs remain immersed in the
hot gas until coalescence. At larger
radii, the geometrically thick flow transitions into a geometrically
thin accretion disk or a lower temperature ambient medium. The
illustrations are not drawn to scale.\label{fig1}}
\end{figure}

\subsection{Circumbinary disk model}\label{S_circumb}

It has been shown that the presence of gas on larger scales in the
aftermath of a gas rich galactic merger may not guarantee an abundant
supply of gas for accretion once a gravitationally bound binary is
formed. This is because the binary torques can evacuate most of the
surrounding gas, preventing in such a way any significant accretion on
either member of the binary \citep{an02,mp05,hayasaki07, hayasaki08,
mm08,cuadra09}. The scenario in which binary torques clear a central
low density region is commonly described in the literature as the {\it
circumbinary disk} (Fig.~\ref{fig1}). In this scenario the orbital
evolution of a gravitationally bound SBHB initially occurs on the
viscous timescale of the surrounding disk, until the binary reaches
the gravitational wave regime. At that point, the rapid loss of
orbital energy and angular momentum through gravitational radiation
cause the binary to detach from the circumbinary disk and to
accelerate towards coalescence. After decoupling, the accretion rate
onto the black holes is expected to diminish and, consequently, the EM
signatures associated with the binary may cease. How early in the
inspiral phase this happens depends on the properties of the
disk. Here we outline the conditions in the circumbinary disk at the
decoupling. We follow \citet{an02}, who note that the viscous rate at
which the inner edge of a Shakura-Sunyaev disk \citep{ss73} moves
inwards is
\begin{equation}
{\dot a}_{\rm visc}=-{3\over 2}\left(h\over r\right)^2\alpha v_K
\end{equation}
where $\alpha$ is the viscosity coefficient, $v_K=\sqrt{GM/a}$, the
Keplerian circular velocity at a semimajor axis $a$, and $h$ is the
half-thickness of the disk at a given radius, $r$. The rate at which
the semimajor axis of a circular binary shrinks is
\begin{equation}
{\dot a}_{\rm GW}=-\frac{64 G^3}{5c^5} \frac{q M^3}{(1+q)^2 a^3}\; .
\end{equation}
Here, $M$ and $q\leq 1$ are the mass and mass ratio of the SBHB.
Setting the two rates equal, while adopting a thin disk aspect ratio
and a commonly used value for $\alpha$ gives
\begin{equation}
a_{\rm decoup} \approx 370\,M \frac{q^{2/5}}{(1+q)^{4/5}}
\left(\frac{h/r}{10^{-2}}\right)^{-4/5}
\left(\frac{\alpha}{0.1}\right)^{-4/5} \,\; .
\end{equation}
We use geometrized units with $G=c\equiv 1$, and so $GM/c^2 = M =
1.48\times 10^{12}\,{\rm cm}\; M_7$ and $M_7 = M\,/(10^7\,M_\odot)$. For an equal
mass binary $a_{\rm decoup}\approx 150\,M$. In reality, the inner edge
will be farther away, because for comparable-mass black holes the disk
is truncated at about double the binary semimajor axis
\citep{al94}. This implies the inner edge of the disk is at $\sim
300M$. We estimate the luminosity and temperature of such a disk using
the Shakura-Sunyaev disk solution. The luminosity of the disk is $L =
\epsilon_r\, \dot{M}c^2$, where we adopt
$\epsilon_r=1 - (r-2)/\sqrt{r(r-3)}$, the maximum efficiency with which
the accretion disk around a Schwarzschild black hole converts the
potential energy into radiative energy at a given radius, r, and
$\dot{M}$ is the accretion rate. Note that at $r = 300M$,
$\epsilon_r\approx 0.0016$ is much less than the radiative efficiency
of the disk that extends all the way to the innermost stable orbit
(ISCO), where $\epsilon_r^{ISCO}\approx 0.057$. To estimate the maximum
mass accretion rate we use Shakura-Sunyaev expression for
half-thickness of the disk in the gas pressure-dominated region
\begin{equation}
\fl h = 1.8\times 10^{12}~{\rm cm}\; \left(\frac{\alpha}{0.1}\right)^{-1/10}
\left(\frac{\dot M}{{\dot M}_E}\right)^{1/5}
\left(\frac{r}{300M}\right)^{21/20}
\left[1-\left(\frac{r}{6M}\right)^{-1/2}\right]^{1/5}\; M_7^{9/10} \;\, ,
%
\end{equation}
where ${\dot M}_E=L_E/(\epsilon_r^{ISCO}c^2)$ and $L_E=1.3\times
10^{45}$~erg~cm$^{-2}$~s$^{-1}\; M_7$ is the Eddington
luminosity. Using $h/r=10^{-2}$ and $r=300\,M$ we calculate the
accretion rate as $\dot{M} = 2.5\;{\dot M}_E$ and consequently, $L =
2.5\;\, (\epsilon_r/\epsilon_r^{ISCO})\;L_E \approx 0.07\;L_E$.
Using the same model we evaluate the temperature of the disk
\begin{equation}
\fl T = 8.3\times10^5\,K\; \left(\frac{\alpha}{0.1}\right)^{-1/5}
\left(\frac{\dot M}{2.5{\dot M}_E}\right)^{2/5}
\left(\frac{r}{300M}\right)^{-9/10}
\left[1-\left(\frac{r}{6M}\right)^{-1/2}\right]^{2/5}\; M_7^{-1/5} \, ,
%
\end{equation}
or, $T = 7.8\times 10^5\,K$. These values imply that geometrically
thin circumbinary disks around comparable mass binaries are expected
to be moderately luminous with spectral energy distribution peaking in
the UV band, as long as the inner region of the disk remains largely
depleted of gas. If gas flow across the ``hole'' region is present (as
illustrated in Fig.~\ref{fig1}), it could feed the circumblackhole
disks and give rise to shocks in which case the coalescing system may
appear more luminous and emit harder radiation. It is worth noting
that in order to maintain accretion disks around individual SBHs for a
longer period leading to the coalescence, the accretion rate between
the circumbinary disk and smaller circumblackhole disks would need to
be carefully balanced. It is not obvious that such a steady state can
be naturally achieved, especially for comparable mass binaries,
because the gas in the vicinity of the binary is subject to strong,
non-axisymmetric binary torques and is not in dynamic
equilibrium. However, the accretion process need not be a steady state
to be luminous, since a relatively small amount of gas in this region
is sufficient to produce a luminous event \citep{krolik10}.

Clearly, the absence of gas in the central region would represent a
fundamental obstacle for the EM searches of coalescing binaries. If
so, even if the circumbinary disk itself is sufficiently luminous to
be observed, the absence of characteristic EM variability would make
it difficult to discern a binary from a ``regular'' active galactic
nucleus (AGN). Nevertheless, several binary candidates discovered so
far have been interpreted in the context of a thin (or moderately
thick) circumbinary accretion disk scenario. One is a well known
binary candidate, the blazar OJ~287, that exhibits outburst activity
in its optical light curve with a period close to 12 years,
interpreted as a signature of the orbital motion
\citep{valtonen08}. It is, however, hard to make inferences about a
population of SBHBs based on one object.  While indications of a
long-term periodicity may also exist in a handful of other objects,
they are generally less pronounced than in case of
OJ~287. An additional observational technique for selection of the
SBHB candidates anchors on the existence of the circumbinary accretion
disk and individual circumblackhole accretion disks. This method
utilizes the emission-lines associated with multiple velocity systems
in the spectrum of a candidate object. It relies on a detection of the
Doppler-shift that arises from the orbital motion of a binary. So far
a handful of SBHB candidates have been selected in this way:
J092712.65+294344.0 \citep{bogdanovic09a, dotti09},
J153636.22+044127.0 \citep{bl09}, J105041.35+345631.3
\citep{shields09}, 4C+22.25 \citep{decarli10}, and J093201.60+031858.7
\citep{barrows10}. A common property shared by all objects is that
their emission line systems exhibit large shifts in the range
$2000-9000\,{\rm km\,s^{-1}}$. If such velocity shifts are interpreted
in the context of a binary model, and if the emission-line systems are
associated with circumblackhole accretion disks around one or both of
the black holes, the expectation is that the orbital motion of the
binary should give rise to a measurable velocity change of the
emission lines on a time-scale of several years. For the first four
objects this basic test was carried out by taking spectra of the
object at different epochs and measuring the change in the position of
the emission line peaks. In all four the measured rate of velocity
shift is very low ($dv/dt\sim{\rm few}\times10\,{\rm
km\,s^{-1}\,yr^{-1}}$) and consistent with zero within the measurement
error bars. This strongly challenges the hypotheses that these
candidate objects are binaries, but the uncertainties in emission
geometry and radiative processes of accretion flows around binaries
currently preclude elimination of the binary model. The lack of
definite observational signatures associated with the gravitationally
bound binaries, as expected in the context of the circumbinary disk
model, leaves open a possibility that accretion flows in vicinity of
the SBHBs may have a different structure.

\subsection{Radiatively inefficient hot gas flow}\label{S_riaf}

It is plausible that if the surrounding gas is sufficiently hot and
tenuous, the binary may find itself engulfed in a radiatively
inefficient, turbulent flow all the way through coalescence. Such
conditions are expected to exist in nuclear regions of some low
luminosity AGN \citep[LLAGNs;][for example]{quataert99, ptak04,
nemmen06, eh09}. If binaries indeed do exist in radiatively
inefficient accretion flows (RIAFs), then the \emph{circumbinary disk} and
\emph{hot gas flow} scenarios effectively bracket the range of
physical situations in which pre-coalescence binaries may be found in
centers of galaxies.  Which scenario prevails depends on the balance
of heating and cooling mechanisms in the accretion flow. In the
circumbinary disk scenario, relatively efficient cooling processes
result in the gas settling into a rotationally-supported,
geometrically thin accretion disk around the binary. The main property
of radiatively inefficient flows is that very little energy generated
by accretion and turbulent stresses is radiated away. Instead, it is
stored as thermal energy in the gas at the level comparable to its
gravitational potential energy, giving rise to a very hot flow
\citep{ichimaru77,rees82,ny94}.  It follows that $T \propto M/R$,
where $M$ is the binary mass and $R$ is the separation from its center
of mass. Since thermal pressure forces within the gas are significant,
the hot accretion flows are expected to be geometrically thick. Also,
given high thermal velocities, the inflow speeds are comparable to the
speed of sound and to the orbital velocity that a test particle would
have at a given radius. This implies that in the hot gas flow, unlike
the circumbinary disk scenario, binary torques are incapable of
creating the central low density region, because the gas ejected by
the binary is replenished on the dynamical time scale. Consequently,
the binary may remain immersed in the gas until coalescence. This may
bode well for the EM searches for coalescing binaries should such hot
gas flows remain sufficiently luminous through the coalescence. We now
discuss the properties of such a plasma and estimate its
characteristic luminosity.

Radiatively inefficient hot gas flows are characterized by lower gas
densities than their radiatively efficient counterparts (such as
accretion disks described in \S~\ref{S_circumb}) and here we use
density as an effective discriminator between the two scenarios. Below
some critical gas density, in RIAF-type flows, the time scale for the 
Coulomb collisions between electrons and ions ($t_{\rm Coulomb}$) becomes
longer than the inflow time of the gas ($t_{\rm inflow}$). This may
result in the formation of a two-temperature flow in which the ion plasma remains at
$T_p\sim 10^{12}\,{\rm K}$ and the electron plasma cools to
temperatures in the range $T_e\sim10^{10}-10^{12}\,{\rm K}$
\citep{ichimaru77,rees82,ny94}.  Above this density limit, the
collisional plasma flow of electrons and ions is fully thermally
coupled and can cool efficiently via electron-emitted radiation,
yielding an evolution more similar to the accretion disk scenario. We
calculate this critical plasma density by setting $t_{\rm Coulomb} =
t_{\rm inflow}$, assuming $T_p = 10^{12}\,{\rm K}$ and $T_e =
10^{10}\,{\rm K}$, and find it to be $\rho_{\rm c} \sim 10^{-11} {\rm
g\, cm^{-3}}$. The characteristic timescales evaluated at the critical
density and in vicinity of the binary are:
\begin{eqnarray}
t_{\rm inflow} = \frac{R_B}{c_s} \approx  0.23\,{\rm hr}
\left(\frac{T_p}{10^{12}\,{\rm K}}\right)^{-3/2} M_7 , \\
t_{\rm Coulomb}  = \frac{1}{n\,\sigma c_s} 
\approx  0.38\,{\rm hr} \left(\frac{\rho_{\rm c}}{10^{-11} {\rm g\, cm^{-3}}}\right)^{-1} 
\left(\frac{T_p}{10^{12}\,{\rm K}}\right)^{-1/2} \left(\frac{T_e}{10^{10}\,{\rm K}}\right)^2 \,.
\end{eqnarray}
The size of the region under consideration, $R_B\approx GM/c_s^2$, is
the Bondi radius of gravitational influence of the coalescing
binary. $n \approx n_p \approx n_e \approx \rho/m_p$ is the number
density of the gas, and $\sigma\approx 0.3 \,Z^2 e^4/ (kT_e)^2$ is the
cross-section for Coulomb scattering of an electron with kinetic
energy $\sim kT_e$ on a more massive ion. $c_s = (\gamma k
T_p/m_p)^{1/2}$ is the speed of sound evaluated assuming the equation
of state of an ideal gas and $\gamma=5/3$, for monoatomic gas. Other
symbols have their usual meaning.
 
The two orbiting SBHs will accrete from a hot, turbulent flow in a
Bondi-like fashion
\begin{equation}
\dot{M}_{\rm B} \approx 0.38\,M_{\odot}\,{\rm yr^{-1}} 
\left(\frac{\rho_{\rm c}}{10^{-11} {\rm g\, cm^{-3}}}\right)
\left(\frac{T_p}{10^{12}\,{\rm K}}\right)^{-3/2} M_7^2 \, , 
\label{eq:bondi}
\end{equation}
where we assumed that the relative velocity between the gas and each
\bh{} is comparable to $c_s$. The luminosities due to bremsstrahlung,
synchrotron, and inverse Compton radiation \citep{rl86} from such flow
can be estimated as
\begin{eqnarray}
 L_{\rm brems}  &\approx &  1\times 10^{44}\,{\rm erg\,s^{-1}} \left(\frac{\rho_{\rm c}}{10^{-11} 
{\rm g\, cm^{-3}}}\right)^2 \left(\frac{T_p}{10^{12}\,{\rm K}}\right)^{-3}
    \left(\frac{T_e}{10^{10}\,{\rm K}}\right)^{1/2}\nonumber\\ 
   & \times & \left[ 1 + 4.4\times\left(\frac{T_e}{10^{10}\,{\rm K}}\right) \right]_{5.4} M_7^6 
\label{eq_Lbrem}
\end{eqnarray}
\begin{eqnarray}
L_{\rm synchro} \approx 2\times10^{36}\,{\rm erg\,s^{-1}} 
\left(\frac{\rho_{\rm c}}{10^{-11} {\rm g\, cm^{-3}}}\right)
\left(\frac{T_p}{10^{12}\,{\rm K}}\right)^{-3} \left(\frac{B}{1G}\right)^2  M_7^6 \label{eq_Ls}\\
L_{\rm IC} \approx 8\times10^{-9}\, L_{\rm soft} 
\left(\frac{\rho_{\rm c}}{10^{-11} {\rm g\, cm^{-3}}}\right)
\left(\frac{T_p}{10^{12}\,{\rm K}}\right)^{-3} \left(\frac{R_{\rm tran}}{10^5 M}\right)^{-2} 
M_7^4\label{eq_Lic}
\end{eqnarray}
where bremsstrahlung luminosity was calculated assuming a thermal
distribution of electrons and the subscript ``5.4'' indicates that the
numerical factor in the square brackets is normalized to 5.4. The
synchrotron luminosity is evaluated for relativistic electrons with
$\beta= v/c \approx 0.3$. In (\ref{eq_Lic}), $L_\mathrm{soft}$
represents a supply of low energy photons transported from the edge of
the RIAF, a distance of $R_\mathrm{tran}$ away. Note that the
luminosity of the synchrotron emission remains below that of the
bremsstrahlung radiation unless the magnetic field strength is close
to the equipartition value, which in our case is $B_{\rm equip}\sim
10^5\,G$. It is plausible that the spinning \bh{s} can amplify the
magnetic fields in their vicinity to nearly the equipartition value
\citep{palenzuela10}. Given the properties of the gas flow in our
model and assuming $B\sim 10^4\,G$, the estimated luminosity of
synchrotron radiation becomes significant, $L_{\rm synchro}\sim
2\times 10^{44}\,{\rm erg\,s^{-1}}$. Similarly, in order for the
inverse Compton luminosity to be significant, a supply of soft, lower
energy photons is required. In estimating $L_{\rm IC} $, we assumed
that this soft photon component is produced at large radii, where
radiative cooling is efficient and the geometrically thick flow
described here transitions into a geometrically thin accretion disk or
a lower temperature ambient medium (see the schematic representation
in Fig.~\ref{fig1}). In the case of Sgr~${\rm A}^\star$ for example,
observations indicate that a radiatively inefficient accretion flow
extends into the ambient medium out to $R_{\rm tran} \sim 10^5\,M$
away from the center~\citep{quataert03}. The estimate for $L_{\rm IC}$
obtained with this value of the transition radius implies that inverse
Compton scattering is a very inefficient process even if a generous
supply of low energy photons is available from the distant ambient
medium, parametrized here in terms of the luminosity $L_{\rm
soft}$. Note however that if synchrotron radiation is boosted by
strong magnetic fields, it could feed the inverse Compton luminosity
of comparable magnitude by providing an immediate source of soft
photons. For comparison, the Eddington luminosity of the system is
$L_E\approx 1.3\times10^{45}\,{\rm erg\,s^{-1}}\,M_7$.

The corresponding cooling timescale of the plasma at the critical
density due to bremsstrahlung radiation (while keeping in mind the
potential importance of the other two mechanisms) is
\begin{eqnarray}
t_{\rm cool} & \sim & 8\,{\rm hr} 
\left( \frac{\rho_{\rm c}}{10^{-11} {\rm g\, cm^{-3}}}\right)^{-1} 
\left(\frac{T_p}{10^{12}\,{\rm K}} \right)
\left(\frac{T_e}{10^{10}\,{\rm K}} \right)^{-1/2} \, .
\end{eqnarray}
Note that $t_{\rm cool} > t_{\rm Coulomb} > t_{\rm inflow}$ implies
that the hot gas plunges into the \bh{s} before it had a chance to
radiatively cool and settle into an accretion disk, as expected in the
case of radiatively inefficient flow.

We evaluated the properties of the hot gas flow at the critical
density, however note that by construction radiatively inefficient
flows reside at densities $<\rho_{\rm c}$ and so the derived
luminosities should be regarded as upper limits. Because $L_{\rm
brems}$, $L_{\rm synchro}$ and $L_{\rm IC}$ are sensitive functions of
density and SBHB mass, it follows that a large fraction of AGNs
associated with this type of accretion flow will likely have low
luminosities. More specifically, the properties of these sources
appear consistent with the group of the LLAGNs observed in the nearby
universe that have been proposed to host radiatively inefficient
accretion flows. \citet{eh09} find that below a bolometric luminosity
of $L\sim 5\times 10^{39}\,{\rm erg\,s^{-1}} M^{2/3}_7$ AGNs
characterized by the broad emission lines (i.e., type 1 AGN) cease to
exist while narrow line (type 2) sources can exist both below and
above this limit in the luminosity range $10^{38} - 10^{42}\,{\rm
erg\,s^{-1}}$ for a central BH mass of $10^7\,\Msun$. According to the
unification model of AGN the absence of the broad emission lines in
type 2 sources can be explained by the observer's orientation with
respect to the toroidal obscuration of the broad line region (BLR) and
central AGN. However, despite the considerable success of the
unification scheme, there is growing evidence that the BLR is actually
missing, and not just hidden, in many LLAGN \citep{ho08}. These
sources are commonly referred to as ``true'' type 2 AGNs. The current
paradigm suggests that the BLR naturally arises
from a clumpy wind coming off an accretion disk rotating around the BH
\citep{emmering92}. In the context of this paradigm, the disappearance
of the BLR at lower luminosities in type 1 sources and absence of it
in true type 2 AGNs signals a change in the properties of the accretion
flow associated with the inner $\sim 1$~pc and can be explained by
the transition from the radiatively efficient to radiatively
inefficient accretion flows. If radiatively inefficient flows around
SBHBs indeed have properties similar to LLAGNs, they would appear as
low to moderate luminosity {\it narrow line} AGN. Note that in this
case the method for selection of binary candidates based on the
detection of the velocity shift or an unusual shape of the broad
emission lines may not be useful. It is also plausible that some
fraction of radiatively inefficient EM counterparts could be as faint
as the accretion flow in the center of our Galaxy \citep{narayan95,
narayan98, quataert03} in which case the EM detection of a binary
would be unlikely even in the local universe.

If indeed relevant to SBHB systems, the properties of the radiatively
inefficient flows could naturally explain the paucity of SBHBs in EM
searches, which have so far been predominantly organized around the
assumption of radiatively efficient accretion flows. Furthermore,
radiatively inefficient accretion flows are characterized by a hard
spectral energy distribution, peaking at $kT_e\sim 100keV - 1MeV$. It
is worth noting that the {\it Swift}-BAT survey of the hard X-ray
sources \citep[for description of the most recent catalogs see][]
{cusumano10,baumgartner10} is uniquely suited for detection of nearby
(z<0.05), moderate-luminosity AGN with such properties, because it is
conducted in the 14$-$195~keV energy band. In this sample
\citet{koss10} find a higher incidence of galaxies with signs of
disruption and higher incidence of AGN and galaxy pairs within 30 kpc separation
compared to a matched control sample of galaxies. This may have
interesting implications for gravitationally bound and coalescing
binaries at much smaller separations, if it can be shown that they
reside in similar environments.

\section{Simulations of black hole coalescence in radiatively inefficient accretion flow}

\begin{figure}[ht]
\center{
\includegraphics[width=0.8\linewidth]{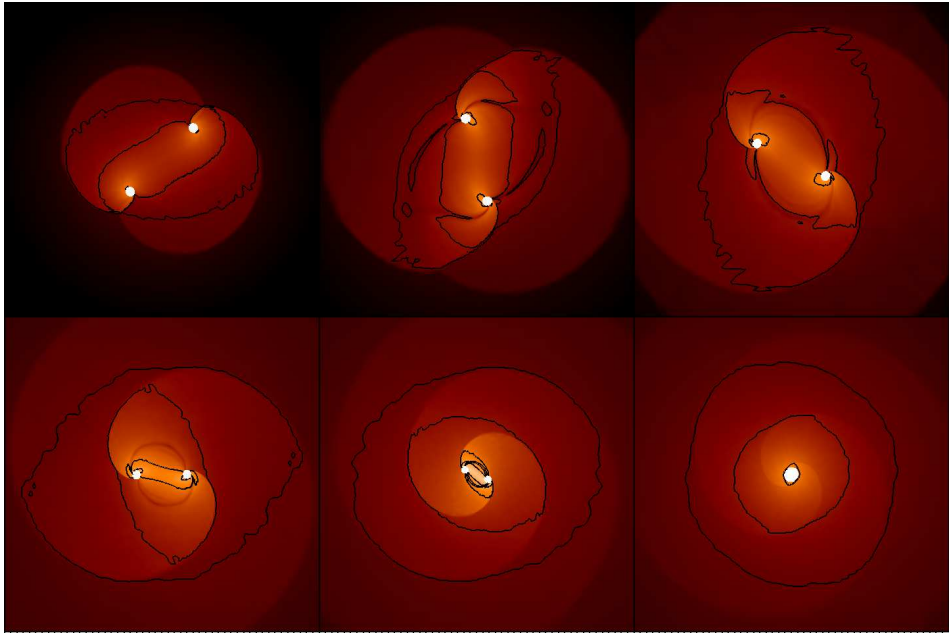}}
  \caption{Snapshots of the rest mass density of the gas, in the
    orbital plane of the binary. The color scale is logarithmic and
    lighter shades mark higher density.\label{fig2}}
\end{figure}

Guided by the analytic expectations described in the previous section
we set out to numerically investigate how properties of the binary
environment correlate with observational signatures associated with
the SBHB coalescence.  This is a work in progress that encompasses a larger
suite of simulations and some of its initial results have been 
described in \citet{bode10}. For the purpose of this proceeding we describe 
one of the subsequently modeled scenarios in order to illustrate the EM signatures 
characteristic for the hot accretion flow scenario. This investigation should be
considered in the context of the relativistic calculations that investigate the 
dynamics of test particles \citep{vanmeter09} and the evolution of EM fields \citep{palenzuela09,ply09,palenzuela10,mosta10} and gas \citep{bode10,farris10} in 
the gravitational potential of a coalescing binary. 

We use fully relativistic numerical hydrodynamics simulation
to follow the interaction of an equal-mass SBH binary with a gaseous
environment through the last five orbits and merger. The simulation
described here has been carried out with the new version of the
\maya{} code of the numerical relativity group at Georgia
Tech~\citep{Bode:2010Maya}.\footnote{\maya{} is a modified version of
the publicly available code \whisky{} developed by the European
Union Network on Sources of Gravitational Radiation \citep[{\tt
http://www.whiskycode.org};][]{baiotti05}.} The data consist of a SBHB
immersed in a uniform gaseous medium with the initial temperature $T_p
\sim 10^{11}$~K. The binary separation at the beginning of the
simulation is $8\,M\approx 10^{-5}\,M_7\,\mathrm{pc}$ and SBHs
have spins that are equal, parallel, and aligned with the orbital
angular momentum. The dimensionless spin parameters are $a = Jc/Gm^2 =
0.6$ and $m = M/2$ is the mass of each black hole. We consider the
scenario in which the total rest mass of the gas in the SBHB vicinity
is much smaller than the BH masses and therefore the dynamics of the
binary is effectively the same as in vacuum. After the initial phase
of relaxation, the gas in the vicinity of the binary orbit settles to
a temperature $T_p \sim 10^{12}$~K, and its thermal velocity is
comparable to the binary orbital speed. The gas is evolved assuming
the equation of state of ideal gas and $\gamma = 5/3$. The only
mechanisms for heating and cooling of gas in our simulation are
adiabatic compression and expansion in the potential of the binary and
we do not account for heating of the gas by the AGNs nor for cooling
by emission of radiation. In order to avoid any spurious transients that may
stem from the imperfections in the initial conditions the binary and gas dynamics 
have been relaxed during the initial $\Delta t\sim 100\,M$ of the simulation. 
We removed this initial phase and do not show it as a part of the results reported
here.\footnote{Description of the numerical setup utilized in our runs can be
found in Bode et al. (2010a).}
 
Fig.~\ref{fig2} shows the snapshots of the rest mass density of the
gas. The distinct features that arise in the gas during binary evolution are the
density wakes that develop behind the inspiraling BHs and a high
density region enclosed in the binary orbit. Early in the inspiral,
the shocks are confined to the wakes expanding outside of the
binary orbit. Later on, the dynamically unstable region between the
two BHs gives rise to another shock region which is swallowed by the
remnant hole at the time of the coalescence. We find that shock
heating of the gas by the binary and subsequent dynamics can give rise
to a characteristic variability of the EM emission signatures which,
if observable, could be uniquely associated with merging binaries in
hot accretion flows.

\begin{figure}[ht]
\center{
\includegraphics[width=0.7\linewidth]{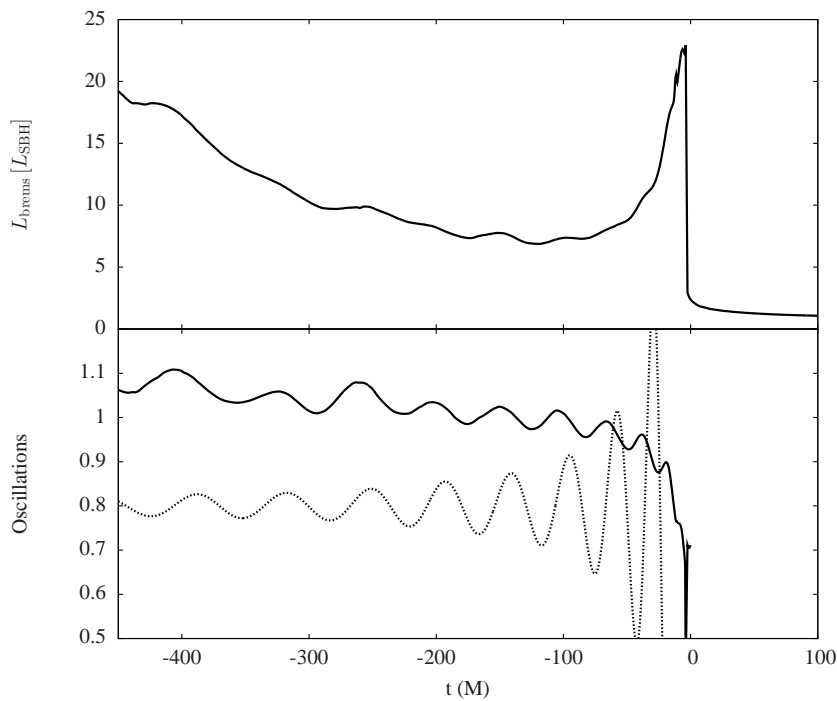}}
\caption{{\it Top:} Bremsstrahlung luminosity as a function of time
normalized by the luminosity of the system in which the binary SBH was
replaced by a single black hole of the same mass. {\it Bottom:}
Comparison of bremsstrahlung and GW variability. Real part of the 
gravitational waveform (dotted line) is arbitrarily scaled and superimposed on the top of the
oscillations in the bremsstrahlung light curve (solid). The latter
are obtained by division of the curve in the top panel by the unbeamed
light curve. Merger occurs at $t\sim0$.\label{fig3}}
\end{figure}

To assess this we evaluate the bremsstrahlung luminosity, $L_{\rm
brems}$, arising from the gas near the binary by integrating the
emissivity of the gas over the volume with radius $20\,M$ but
excluding the volumes inside the apparent horizons of the two BHs. Top
panel of Fig.~\ref{fig3} shows the light curve calculated for this run
where $t = 0$ marks the time of the merger. The luminosity in
Fig~\ref{fig3} was normalized to the luminosity of the gas around a
single BH with the mass equal to that of the binary. The luminosity of a 
single BH system can be parametrized in terms of the BH mass and 
the gas properties and its upper
limit is set by the value shown in (\ref{eq_Lbrem}). Fig~\ref{fig3}
illustrates that the shocks driven by the orbiting binary provide an
extra source of heating to the gas, causing it to be more luminous
than in the single BH scenario. At about $t \sim 100\,M \approx
5\times10^3\,M_7\,{\rm s}$ before the merger, when the binary enters
the final plunge, the luminosity curve exhibits a broad flare.  The
flare reaches $L_\mathrm{brems} \approx 20 L_{\rm SBH}$ before a
sudden drop that occurs soon after the BHs have merged. A sudden
decrease can be attributed to the disappearance of the dynamic region
of high emissivity between the two BHs, which is rapidly swallowed by
the BHs in the process of coalescence. Given a large magnitude of the
luminosity drop, it is possible that a source that was initially
visible on the sky may disappear shortly after coalescence, making
this its characteristic signature.

The bottom panel of Fig.~\ref{fig3} shows the quasi-periodic
oscillations in the bremsstrahlung light curve (solid line) that arise
due to the relativistic beaming and Doppler boosting of the light
emitted by the high density wakes that develop behind the inspiraling
BHs. To emphasize the oscillations we divided the light curve in the
top panel by the shape of the unbeamed light curve.\footnote{In
calculation of the luminosity of the beamed light we neglected
relativistic bending of photon trajectories and gravitational redshift
of photons in the potential well of the binary.} The amplitude of the
variations in luminosity is approximately $10\%$ between subsequent
peaks and troughs, as seen by a fiducial observer placed in the plane
of the binary at infinity. In this way, the observer is placed
directly in the path of the sweeping emission beams associated with
the two BHs and can sample both the minimum and maximum luminosity of
the system, depending on its orbital phase. The amplitude of the
variability sensitively depends on the distribution gas around the
SBHB and we find it to be smaller in this run compared to the
previously reported scenario where the SBHB was immersed into a finite
size ``cloud'' \citep[the amplitude in that case was $\sim
2$;][]{bode10}. 

The dashed line in the bottom panel of Fig~\ref{fig3} shows the
GW variability in form of the arbitrarily scaled and shifted real
part of the waveform.  In this model the frequency of oscillations 
observed from the bremsstrahlung's relativistically beamed light 
is directly tied to the orbital dynamics of the binary and thus is 
also correlated with the frequency of the GWs in this run. As 
mentioned before, detection of correlated
EM+GW oscillations from the same object would be a smoking gun for a
SBHB system on the way to coalescence and would directly link a
detected GW source to its EM counterpart. It is important to point out that
the quasi-periodic EM variability is only expected to arise in systems
where the two BHs can form a stable set of symmetric wakes in the
gas. However, as most SBHB systems in the universe are expected to be
unequal-mass binaries, and thus have some inherent asymmetry, the
likelihood of observing such correlated oscillations therefore depends
on the extent to which asymmetries can modify the variability. It is
also worth noting that the discussed EM variability stands a chance of
being seen by a distant observer if the amplitude of the
quasi-periodic variability stands above the intrinsic variability of an
AGN.  At the $\sim 10\%$ level the amplitude of the quasi-periodic
oscillations is comparable to the natural variability of the AGNs and
unless quasi-periodic oscillations can be more pronounced, they will 
remain hidden by the AGN flickering. An
additional requirement is that the photons emitted in the vicinity of
the BHs are not absorbed by the surrounding medium or reprocessed in
such a way that the variability signature is lost. If the cloud is
optically thick to emitted radiation, the characteristic variability
will most likely be ``erased'' and dephased during the reprocessing of
photons by the intervening medium. In this case a gradual rise and drop-off in
bremsstrahlung luminosity may be more robust signatures of binary
coalescence as they are still expected to arise even in the case of the
opaque ambient medium.

\section{Summary and future prospects}

In this paper we summarized the key properties of accretion flows
around coalescing supermassive black holes in the context of the two
different models: circumbinary accretion disk and a hot, radiatively
inefficient accretion flow. The models effectively explore two
opposite ends of the parameter space, in terms of the thermodynamic
properties of the binary environment, and bracket a range of physical
scenarios in which coalescing SBHBs can be found. Both classes of
models predict luminous (detectable) or underluminous (undetectable)
outcomes, depending on the details of a particular model. Tauntingly,
astrophysical models currently cannot offer a definitive answer about
the EM signatures associated with coalescences but do offer a variety
of outcomes for consideration.

Guided by the hot, radiatively inefficient accretion flow model, we
carried out a suite of exploratory simulations of the final stages of
binary evolution, where we initially consider only equal mass,
spinning SBHBs. Here we highlighted the key results from one such run,
which shows that correlated EM-GW variability can arise in merging
binary systems immersed in hot gas flows due to the effects of
relativistic beaming and Doppler boosting modulated by the binary
orbital motion. While some degree of quasi-periodic variability is
present in all cases we explored so far, where SBHs form a stable set
of wakes from inspiral through the plunge, there is an indication
that this signature may not be preserved and observable in general. 
In this case,
additional signatures may be sought for in searches for EM
counterparts. We also find that our modeled light curves exhibit a
gradual rise, arriving at a peak at the time of coalescence, followed
by a sudden drop-off. If present, these two features are sufficiently
robust to allow identification of an EM counterpart to a GW
source. The remaining question is whether EM signatures of
coalescences in RIAF type flows will be sufficiently luminous to be
detected. Optimistically, the most luminous solutions physically
allowed in the context of this model imply that most massive binaries
detectable in the LISA band may be identified in X-ray searches out to
$z\approx 0.1 - 1$ (assuming an IXO or EXIST-like X-ray
mission). Pessimistically, if the properties of the SBHB environments
are similar to the hot and underluminous accretion flow in the center
of our Galaxy, detection of EM counterparts will be unlikely. More
moderately, if their emission properties are similar to the observed
LLAGN hypothesized to host RIAFs, then EM counterparts may be
observable in the local universe.

Given the extent of the parameter space involved in coalescing SBHBs
interacting with gas, more follow-up work is needed. As noted before,
most of the SBH binaries in the universe are expected to involve
unequal masses and general spin orientations, and it is thus important
to further explore the parameter space. Similarly, because the
thermodynamic properties of the surrounding gas can significantly
influence the properties of EM signals, in the future we will also
consider the circumbinary accretion disk scenario. An improvement of
current numerical models can be achieved by further inclusion of
relevant physics, such as magnetic fields and radiative transfer,
which will allow more robust predictions about EM counterparts to be
made. Indeed, given recent and complementary work by several different
groups, numerical relativity has decisively stepped into the
``astrophysical regime'', bringing new levels of complexity to
numerical relativistic models but also the opportunity for new
insights into the so far secret life of SBH binaries in the final
stages of their evolution.

\ack Support for Bogdanovi\'c provided by NASA through Einstein
Postdoctoral Fellowship Award PF9-00061 issued by the Chandra X-ray
Observatory Center, which is operated by the Smithsonian Astrophysical
Observatory for and on behalf of the NASA under contract
NAS8-03060. This work was supported in part by the NSF grants 0653443,
0855892, 0914553, 0941417, 0903973, 0955825. Computations described in
this paper were carried out under Teragrid allocation TG-MCA08X009.

\end{document}

%% file: defs.tex
\def\newacronym#1#2#3{\gdef#1{#3 (#2)\gdef#1{#2}}}

\newacronym{\mpm}{MPM}{moving puncture method}

\newacronym{\cra}{CRA}{Center for Relativistic Astrophysics}

\newacronym{\ghm}{GHM}{generalized harmonic method}
\newacronym{\ma}{MA}{Major Activity}
\newacronym{\hpc}{HPC}{high performance computing}
\newacronym{\cgwp}{CGWP}{Center for Gravitational Wave Physics}
\newacronym{\tat}{TAT}{Theoretische Astrophysik T\"ubingen}
\newacronym{\aei}{AEI}{Albert-Einstein-Institute (Potsdam)}
\newacronym{\jena}{Jena}{Friedrich-Schiller-University Jena}
\newacronym{\gsfc}{GSFC}{Goddard Space Flight Center}
\newacronym{\ornl}{ORNL}{Oak Ridge National Laboratory}
\newacronym{\lisa}{LISA}{Laser Interferometer Space Antenna}
\newacronym{\doe}{DOE}{Department of Energy}
\newacronym{\esa}{ESA}{European Space Agency}
\newacronym{\ligo}{LIGO}{Laser Interferometer Gravitational-wave
  Observatory} \newacronym{\lsc}{LSC}{LIGO Scientific Collaboration}
\newacronym{\Caltech}{Caltech}{California Institute of Technology}
\newacronym{\PFC}{PFC}{Physics Frontier Center}
\newacronym{\pfc}{PFC}{Physics Frontier Center}
\newacronym{\EOD}{EOD}{Education, Outreach and Diversity}
\newacronym{\MIT}{MIT}{Massachusetts Institute of Technology}
\newacronym{\sph}{SPH}{smooth particle hydrodynamics}
\newacronym{\amr}{AMR}{adaptive mesh refinements}
\newacronym{\tsi}{TSI}{Terascale Supernova Initiative}
\newacronym{\wmap}{WMAP}{the Wilkinson Microwave Anisotropy Probe}
\newacronym{\decigo}{DECIGO}{the Deci-Hertz Interferometric
  Gravitational-wave Observatory} 
  \newacronym{\cmbr}{CMBR}{cosmic
  microwave background} \newacronym{\ibbh}{IBBH}{intermediate binary
  black hole} \newacronym{\bdj}{BDJ}{Brans-Dicke-Jordan}
\newacronym{\bbo}{BBO}{Big Bang Observer}
\newacronym{\decigo}{DECIGO}{Deci-Hertz Gravitational-Wave
  Observatory} 
\newacronym{\cgwa}{CGWA}{University of Texas
  (Brownsville) Center for Gravitational Wave Astronomy}
\newacronym{\wiser}{WISER}{Women In Science and Engineering Research}
\newacronym{\mure}{MURE}{Minority Undergraduate Research Experience}
\newacronym{\igc}{IGC}{Institute for Gravitation and the Cosmos}
\newacronym{\sfb}{SFB/TR7}{Sonderforschungsbereich Transgerio 7,
  Gravitational Wave Astronomy Methods, Sources \& Observations}
\newacronym{\grb}{GRB}{Gamma-ray burst}
\newacronym{\grbs}{GRBs}{Gamma-ray bursts}
\newacronym{\csats}{CSATS}{Center for Science and The Schools}
\newacronym{\STEM}{STEM}{Science, Technology, Engineering and Mathematics}
\newacronym{\aaas}{AAAS}{American Association for the Advancement of Science}
\newacronym{\nlp}{NLP}{Natural Language Processing}
\newacronym{\heasarc}{HEASARC}{High Energy Astrophysics Science Archive Research Center}
\newacronym{\psu}{PSU}{Penn State University}
\newacronym{\gr}{GR}{general relativity}
\newacronym{\EM}{EM}{electromagnetic}
\newacronym{\gw}{GW}{gravitational wave}
\newacronym{\utb}{UTB}{University of Texas (Brownsville)}
\newacronym{\agns}{AGN}{active galactic nuclei}
\newacronym{\adaf}{ADAF}{advection-dominated accretion flow}
\newacronym{\llagns}{LLAGN}{low luminosity AGN}
\newacronym{\FOV}{FOV}{field of view}

\def\isco#1{inner-most stable orbit#1 
  (ISCO#1)\gdef\isco{ISCO}}
\def\emri#1{extreme mass-ratio inspiral#1 
  (EMRI#1)\gdef\emri{EMRI}}
\def\grb#1{Gamma-Ray Burst#1
  (GRB#1)\gdef\grb{GRB}}
\def\da#1{data analysis#1
  (DA#1)\gdef\da{DA}}
\def\nr#1{numerical relativity#1
  (NR#1)\gdef\nr{NR}}
\def\pnw#1{post-Newtonian#1
  (PN#1)\gdef\pnw{PN}}
\def\bh#1{black hole#1
  (BH#1)\gdef\bh{BH}}
\def\bbh#1{binary black hole#1
  (BBH#1)\gdef\bbh{BBH}}
\def\imbh#1{intermediate mass black hole#1
  (IMBH#1)\gdef\imbh{IMBH}}
\def\smbh#1{supermassive black hole#1
  (SMBH#1)\gdef\smbh{SMBH}}
\def\ulx#1{ultra-luminous x-ray source#1 (ULX#1)\gdef\ulx{ULX}}
\def\lmxbs{low-mass x-ray binaries
  (LMXBs)\gdef\lmxbs{LMXBs}\gdef\lmxb{LMXB}} 
\def\lmxb{low-mass x-ray binary
  (LMXB)\gdef\lmxbs{LMXBs}\gdef\lmxb{LMXB}} 
\def\emrb{extreme mass-ratio binary
  (EMRB)\gdef\emrb{(EMRB}\gdef\emrbs{EMRBs}} 
\def\emrbs{extreme mass-ratio binaries
  (EMRBs)\gdef\emrb{(EMRB}\gdef\emrbs{EMRBs}} 
\def\ns#1{neutron star#1
  (NS#1)\gdef\ns{NS}}
\def\whd#1{white dwarf#1
  (WD#1)\gdef\whd{WD}}
\def\qnm#1{quasi-normal mode#1
  (QNM#1)\gdef\qnm{QNM}}
\def\bhns#1{black hole -- neutron star#1 
  (BHNS#1)\gdef\bhns{BHNS}}
\def\nsns#1{neutron star -- neutron star#1 
  (NSNS#1)\gdef\nsns{NSNS}}
\def\ahz#1{apparent horizon#1 (AH#1)\gdef\ahz{AH}}
\def\tov#1{Tolman-Oppenheimer-Volkoff#1 (TOV#1)\gdef\tov{TOV}}

\newcommand{\Msun}{\mbox{$M_\odot$}}

\newcommand{\siml}{\lower4pt \hbox{$\buildrel < \over \sim$}}
\newcommand{\simg}{\lower4pt \hbox{$\buildrel > \over \sim$}}

\def\simlt{\mathrel{\hbox{\rlap{\hbox{\lower4pt\hbox{$\sim$}}}\hbox{$<$}}}}
\def\simgt{\mathrel{\hbox{\rlap{\hbox{\lower4pt\hbox{$\sim$}}}\hbox{$>$}}}}

\def\ltsima{$\; \buildrel < \over \sim \;$}
\def\ltsim{\lower.5ex\hbox{\ltsima}}
\def\gtsima{$\; \buildrel > \over \sim \;$}
\def\gtsim{\lower.5ex\hbox{\gtsima}}

\def\whisky#1{\textsc{Whisky}#1}
\def\maya#1{\textsc{Maya}#1}